\documentclass[aps,pra,reprint,showpacs,notitlepage,superscriptaddress,twocolumn,floatfix]{revtex4-2}
\usepackage{amssymb}
\usepackage{mathrsfs}
\usepackage{amsfonts}
\usepackage{graphicx}
\usepackage{amsmath}
\usepackage{dcolumn}
\usepackage{color}
\usepackage{subfigure}
\usepackage{epsfig}
\usepackage{soul}
\usepackage{float}
\usepackage[colorlinks,linkcolor=blue,citecolor=blue,hyperindex,bookmarks=false,pdfstartview=FitH]{hyperref}
\usepackage{CJKutf8}
\usepackage{xcolor}
\usepackage[normalem]{ulem}
\providecommand{\U}[1]{\protect\rule{.1in}{.1in}}

\newcommand{\figpanel}[2]{\hyperref[#1]{\ref*{#1}(#2)}}

\begin{document}
\begin{CJK*}{UTF8}{gbsn}

\title{Flexible Control of Temporal Reflection and Refraction via Joint Parameter Modulation}

\author{Liu-Shan Zhang}
\affiliation{College of Physics and Electronic Engineering, Hainan Normal University, Haikou 571158, China}
\author{Cheng-Qi Wu}
\affiliation{College of Physics and Electronic Engineering, Hainan Normal University, Haikou 571158, China}
\author{Dong Yan}
\affiliation{College of Physics and Electronic Engineering, Hainan Normal University, Haikou 571158, China}
\author{Hong Yang}
\affiliation{College of Physics and Electronic Engineering, Hainan Normal University, Haikou 571158, China}

\author{Han-Xiao Zhang}
\email{hxzhang@hainnu.edu.cn}
\affiliation{College of Physics and Electronic Engineering, Hainan Normal University, Haikou 571158, China}
\author{Chu-Hui Fan}
\email{mickvan@163.com}
\affiliation{School of Physics and Optoelectronics, Xiangtan University, Hunan 411105, China}

\date{\today}

\begin{abstract}
Achieving the flexible control of wave propagation is a central goal in the design of optical devices. In this work, our analytical results and numerical simulations show that in both the two-legged model and SSH model, joint modulation of coupling strengths and phases enables independent control of the probability partition and the propagation opening angles in the space-time map. We hope these results could provide a theoretical foundation for the design of programmable time-modulated optical devices.
\end{abstract}

\maketitle
\section{Introduction}
\label{section1}
Precise and flexible control of light propagation has underpinned major advances in optical communication, imaging, sensing, and information processing\cite{Reed2010, Sun2016, Liu2025, Thureja2022}. As optical technologies entering the quantum regime, control over the propagation and internal degrees of freedom of light has also played key roles in quantum communication, quantum information processing, quantum metrology, and quantum networks\cite{Wang2025A,Karpinski2021}. Traditionally, light propagation are controlled by optical optical devices based on carefully engineered spatial structure, such as lenses, prisms, diffraction gratings, beam splitters, and waveguides\cite{Li2020,Amanti2024}. It has long been recognized that, at the level of the wave equation, spatial and temporal variations exhibit a formal analogy\cite{Patera2018, Mendonca2000}, motivating the extension of interface-based wave control from space to time. A temporal interface is formed when the parameters of a spatially homogeneous system undergo an abrupt change in time\cite{Morgenthaler1958,Nefedkin2026}. Unlike a stationary spatial interface, which breaks spatial translational symmetry and generally changes the wave vector while preserving the frequency, a spatially homogeneous temporal interface preserves the wave vector while allowing the frequency or energy band to change\cite{Hayran2022,Caloz2020,Mendonca2002,Zhang2021Temporal}.

At a temporal interface, the wave can split into temporally refracted and reflected components~\cite{Xiao2014,ZuritaSanchez2009,Plansinis2015,Ramaccia2020}. Although the theoretical foundations of temporal refraction and reflection were established decades ago, their experimental realization remained challenging because it requires a sufficient change in the system parameters over the time scale comparable to the relevant wave period\cite{Dong2024,Moussa2023,Vezzoli2018,Guo2026}. Recent advances in ultrafast modulation\cite{Lustig2021Femtosecond}, time-varying metamaterials, and synthetic-lattice technologies have enabled temporal refraction and reflection to be observed in epsilon-near-zero media\cite{Zhou2020}, microwave systems\cite{Jones2024}, synthetic-frequency lattices\cite{Long2023}, and ultracold-atom momentum lattices\cite{Dong2024}. Moreover, beyond a single temporal interface, periodic temporal modulation can give rise to photonic time crystals\cite{Zeng2017}, which exhibit Floquet band structures, momentum band gaps, and wave amplification\cite{Galiffi2022,Wang2023,Wang2025B,Asgari2024,Huang2026Vortex}. These developments have established temporal modulation as a promising approach to control wave propagation\cite{ZuritaSanchez2009}, scattering\cite{Moussa2023}, and frequency conversion\cite{Galiffi2023,Galiffi2025,Huang2025Plasma}.
 
In lattice systems, temporal reflection and refraction can be influenced by changes in coupling strengths\cite{Long2023,Xu2025,Li2026} and phases\cite{Dong2024,Liu2026}. For example, varying the coupling phase can affect both the dispersion relation and the eigenstate structure of the post-interface system. Consequently, it may modify the group velocities, and thus the opening angle between the propagation wave packets, as well as their overlap with the incident state\cite{Jiang2026}, leading to a simultaneous change in the propagation geometry and the temporal-reflection and refraction probabilities\cite{Dong2024}. Hence, systematic and independent control of probability distribution and propagation geometry remains as a challenge. In previous work \cite{Liu2026}, people have demonstrated that in the two-legged model, the probabilities can be modified without changing the propagation angles by tuning the coupling phases. In this work, we further show that both the redistribution probabilities and the corresponding opening angle can be independently controlled via the joint coupling strength and phase modulation, in both the two-legged or SSH model. Our results provide a flexible theoretical framework for the future design of time-modulated optical devices, with potential applications in quantum photonic information processing.

\section{Two-legged Ladder Model}
\label{section2}

\begin{figure}[htbp]
\centering
\vspace{0pt}
\includegraphics[width=0.5\textwidth]{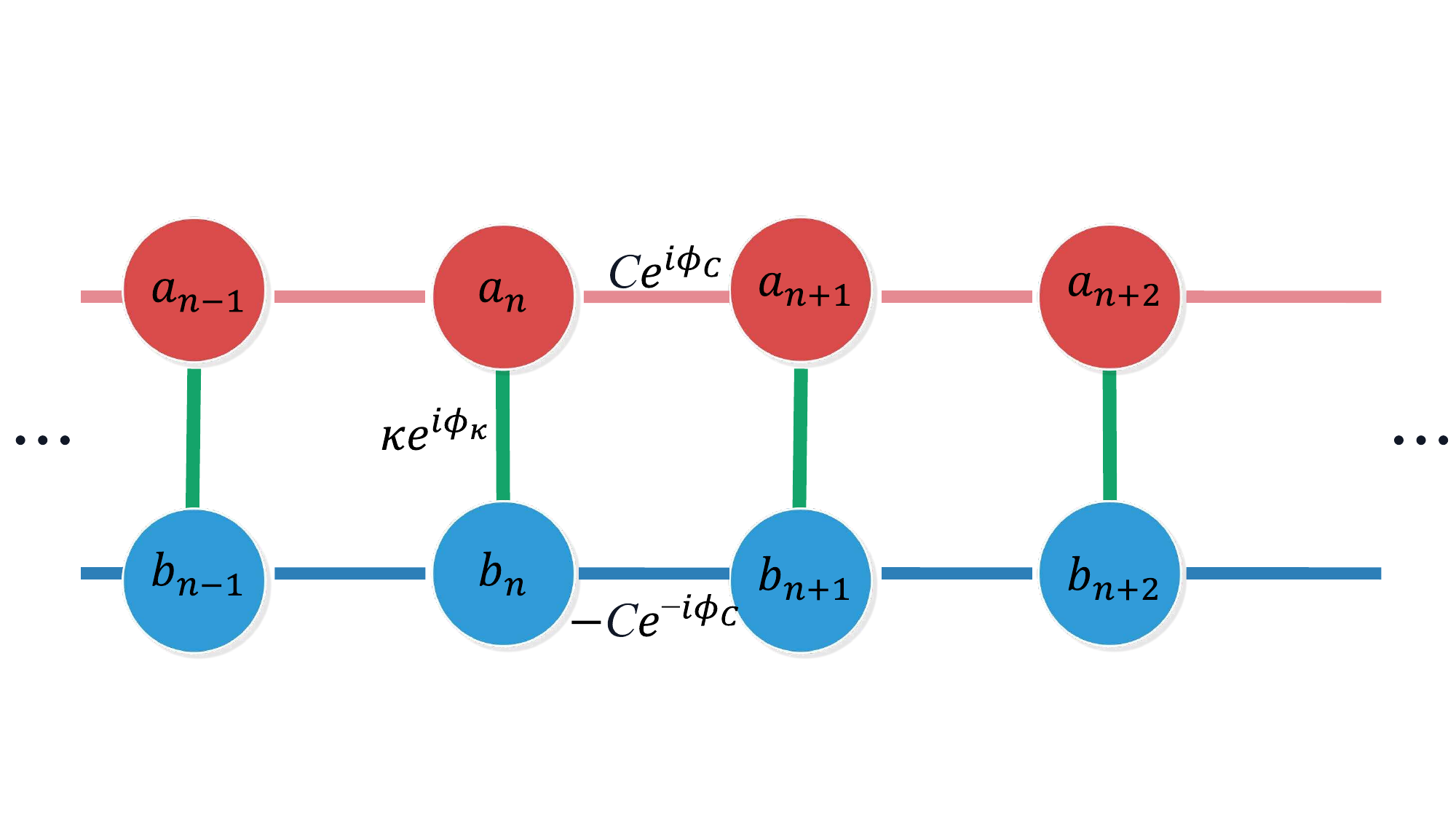}
\caption{Schematic of the two-legged ladder model. The upper (red) and lower (blue) legs have nearest-neighbor hopping amplitudes $Ce^{i\phi_C}$ and $-Ce^{-i\phi_C}$, respectively. Sites on the two legs are coupled by an inter-leg hopping $\kappa e^{i\phi_\kappa}$.}
\label{FIG1}
\end{figure}

In the two-legged ladder model, an abrupt change in either the coupling strength $C$ or the coupling phase $\phi_{\kappa}$ at a temporal boundary can cause the initial eigenstate wave packet to split into temporal refracted and reflected components. However, when only the coupling strength is adjusted, the reflection/refraction probabilities and the corresponding group velocities are generally changed simultaneously. As a result, it is difficult to tune the propagation angle while keeping the probability partition unchanged, or conversely, to tune the probability partition while preserving the propagation angles, by adjusting the coupling strength alone. On the other hand, modulation of the coupling phase $\phi_{\kappa}$ can redistribute the temporal reflection and refraction probabilities without modifying the band dispersion and the group velocities. Therefore, phase modulation alone provides flexible control over the intensity partition, but cannot tune the propagation angles. Since the coupling strength and coupling phase affect different aspects of the temporal-boundary dynamics, a natural strategy is to combine them: one parameter can be used to set the desired propagation geometry, while the other compensates for the accompanying change in the eigenstate projection. This provides a route toward independent control of the temporal reflection/refraction probability ratio and the space-time propagation angles within an accessible parameter range.
As illustrated in Fig.~\ref{FIG1}, the two legs are coupled by an inter-leg hopping amplitude $\kappa e^{i\phi_{\kappa}}$, while the intra-leg hopping amplitudes along the two legs are given by $Ce^{i\phi_C}$ and $-Ce^{-i\phi_C}$, respectively. The corresponding Hamiltonian reads
\begin{align}
H=\sum_n\big[
&Ce^{i\phi_C}a_n^{\dagger}a_{n+1}
-Ce^{-i\phi_C}b_n^{\dagger}b_{n+1} \notag\\
&+\kappa e^{i\phi_{\kappa}}a_n^{\dagger}b_n
\big]+\mathrm{H.c.},
\end{align}
which yields the Hamiltonian in the momentum space

\begin{equation}
H(k)=\begin{pmatrix}2C\cos(k+\phi_C)
&\kappa e^{i\phi_{\kappa}}\\\kappa e^{-i\phi_{\kappa}}
&-2C\cos(k-\phi_C)\end{pmatrix}.
\end{equation}
By introducing
\begin{equation}
\begin{aligned}
d_0(k)&=-2C\sin k\sin\phi_C, \\
d_z(k)&=2C\cos k\cos\phi_C,
\end{aligned}
\end{equation}
the eigenenergies can be written as
\begin{equation}
E_{\pm}(k)=d_0(k)\pm\Omega(k),
\label{eigen_energy}
\end{equation}
where
\begin{equation}
\Omega(k)=\sqrt{d_z^2(k)+\kappa^2},
\end{equation}
and the corresponding eigenstates are
\begin{equation}
\left|\psi_{\pm}(k)\right\rangle
=\frac{1}{N_{\pm}}
\begin{pmatrix}
d_z(k)\pm\Omega(k)\\
\kappa e^{-i\phi_{\kappa}}
\end{pmatrix},
\label{eq:eigenstate_ladder}
\end{equation}
where $N_{\pm}$ are the normalization constants.

We assume that the initial wave packet $\psi_{1+}$ is prepared in the upper-band eigenstate centered at $k_0$. After the temporal boundary, the temporal refraction and reflection probabilities are determined by the projections of the initial state onto the post-boundary eigenstates

\begin{equation}
\begin{aligned}
T &= \left|\left\langle \psi_{2+}(k_0)\middle|\psi_{1+}(k_0)\right\rangle\right|^2, \\
R &= \left|\left\langle \psi_{2-}(k_0)\middle|\psi_{1+}(k_0)\right\rangle\right|^2,
\end{aligned}
\end{equation}
and the corresponding group velocities are determined by
\begin{equation}
v_{\pm}(k)=\frac{\partial E_{\pm}(k)}{\partial k}.
\label{eq:group_velocity_ladder}
\end{equation}
The probabilities $T$ and $R$ characterize the intensity partition between the two temporal channels, whereas the group velocities determine the propagation angles in the space-time map.

To demonstrate how the propagation angle can be tuned while compensating the probability partition, we first consider the case $\phi_C=0$. The intra-leg coupling strength before the temporal boundary $t_{c}$ is defined as $C_1$, and the inter-leg coupling phase is $\phi_{\kappa 1}$. At the temporal boundary $t_c$, the intra-leg coupling strength and the inter-leg coupling phase are changed by $\Delta C$ and $\Delta \phi_{\kappa}$, respectively. For convenience, we define
\begin{equation}
\alpha_1=2C_1\cos k_0,\qquad
\alpha_2=2(C_1+\Delta C)\cos k_0,
\end{equation}
and
\begin{equation}
\beta_j=\sqrt{\alpha_j^2+\kappa^2},\qquad j=1,2.
\end{equation}
In this case, the temporal refraction probability and the corresponding group velocity are given by
\begin{equation}
T=\frac{1}{2}\left[
1+\frac{\alpha_1 \alpha_2+\kappa^2\cos(\Delta\phi_{\kappa})}{\beta_1 \beta_2}
\right],
\label{eq:T_ladder_comp}
\end{equation}
and
\begin{equation}
v_{+}(C_1+\Delta C)=
-\frac{4(C_1+\Delta C)^2\sin k_0\cos k_0}{\beta_2}.
\label{eq:vplus_ladder_comp}
\end{equation}
Similarly, the temporal reflection probability and the corresponding group velocity are
\begin{equation}
R=\frac{1}{2}\left[
1-\frac{\alpha_1 \alpha_2+\kappa^2\cos(\Delta\phi_{\kappa})}{\beta_1 \beta_2}
\right],
\label{eq:R_ladder_comp}
\end{equation}
and
\begin{equation}
v_{-}(C_1+\Delta C)= -v_{+} (C_1+\Delta C).
\label{eq:vminus_ladder_comp}
\end{equation}

%%%%%%%%%%%%%%%%%%%%%%%%%%%%%%%%%%%%%%%%%%%%%%%%%%

\begin{figure*}[t]
\centering
\vspace{0pt}
\includegraphics[width=0.85\textwidth]{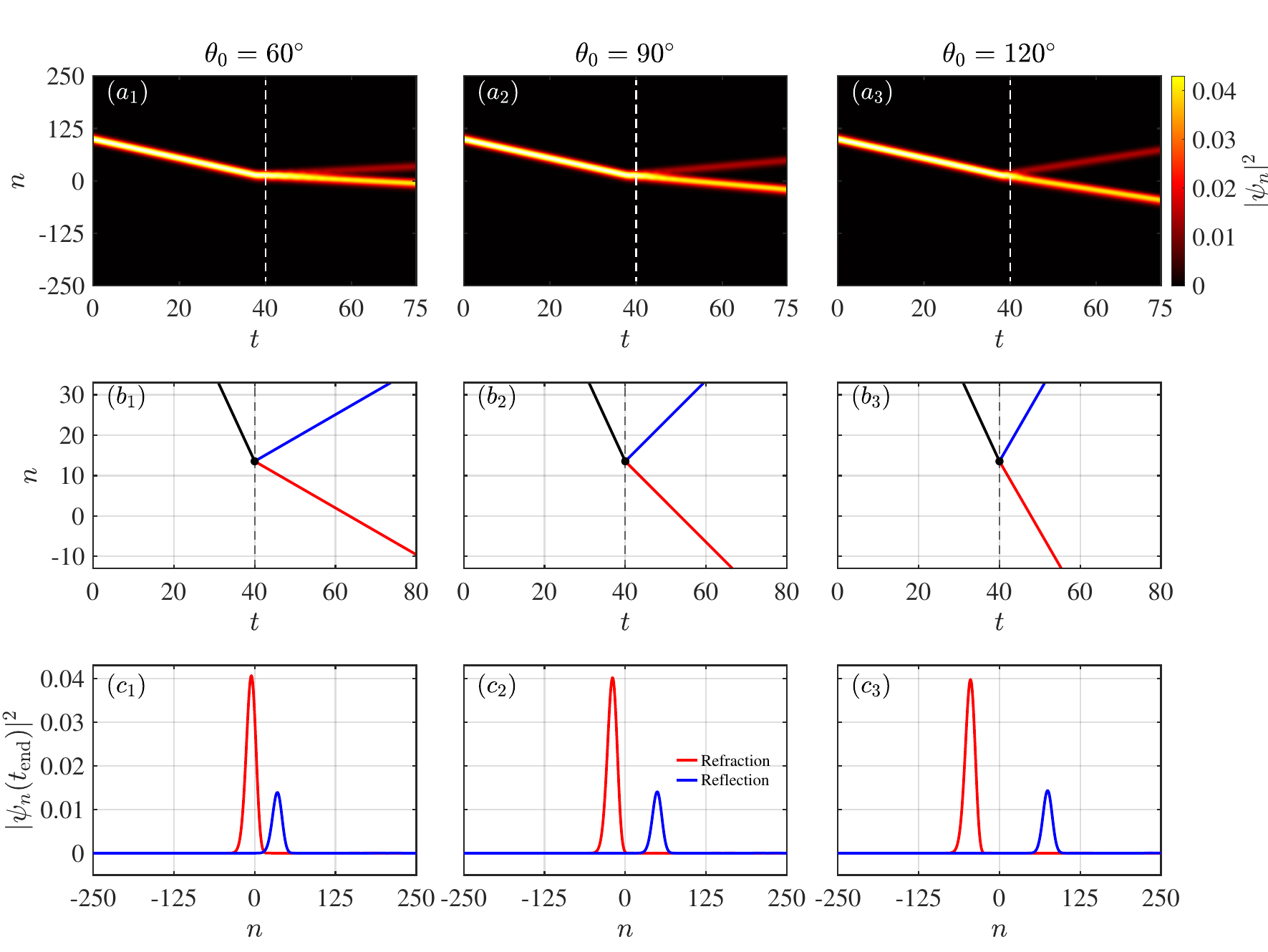}
\caption{Control of the propagation opening angle at a fixed
temporal refraction--reflection probability partition
$T_{0}:R_{0}=0.75:0.25$. The three columns correspond to prescribed
opening angles $\theta_{0}=60^\circ$, $90^\circ$, and $120^\circ$,
respectively. (a$_1$)--(a$_3$) Space--time intensity
distributions. The vertical
dashed lines mark the temporal boundary at $t_c=40$.
(b$_1$)--(b$_3$) Trajectories of the incident and outgoing
wave-packet centers, extracted from the intensity maxima.
The spatial and temporal axes are plotted with equal scales
to display the prescribed opening angles. (c$_1$)--(c$_3$)
Spatial wave-packet intensity profiles at $t_{end}=75$. The initial parameters are set as \(C_1=1.5\), \(\kappa=1\), and $\phi_{\kappa 1}=0.2\pi$. 
}
\label{FIG2}
\end{figure*}

Now let's demonstrate how the preset propagation angle can be achieved while maintaining the target temporal refraction--reflection probability partition via the joint modulation of the intra-leg coupling strength $C$ and the inter-leg coupling phase $\phi_{\kappa}$. Since the inter-leg coupling phase does not enter the band dispersion, the post-boundary coupling strength $C_2 = C_1 + \Delta C$ can first be selected to produce the target propagation opening angle $\theta_0$. With the normalized space--time coordinates, the propagation opening angle is given by
\begin{equation}
\theta_0 = 2\arctan\!\left(\left|v_{\pm}(C_2)\right|\right).
\label{eq:opening_angle}
\end{equation}
For convenience in the following discussion, we define the group velocity associated with the target opening angle as
\begin{equation}
V_{\theta_0} = \tan\!\left(\frac{\theta_0}{2}\right)
= \left|v_{\pm}(C_2)\right|,
\label{eq:target_velocity}
\end{equation}
then the required coupling strength $C_2$ for the preset propagation opening angle can be obtained by inverting the
group-velocity relation, yields
\begin{equation}
\begin{aligned}
C_2(\theta_0)&
= \Bigg\{ \frac{1}{8\sin^2 k_0\cos^2 k_0}
\Bigg[ V_{\theta_0}^2\cos^2 k_0 \\
&+ \sqrt{ V_{\theta_0}^4\cos^4 k_0
+ 4V_{\theta_0}^2\kappa^2 \sin^2 k_0\cos^2 k_0 }
\Bigg] \Bigg\}^{1/2}.
\end{aligned}
\label{eq:C_target}
\end{equation}

Now the desired propagation angle has been achieved by the fine-tuned coupling strength $C_2$. However, the temporal refraction-reflection probability partition also deviates from its preset value because the projections of the initial state onto the post-boundary eigenstates are simultaneously changed. One possible way to compensate this deviation is to appropriately tune the inter-leg coupling phase. Note that the target temporal refraction and reflection probabilities can be written as

\begin{figure}[t]
\centering
\vspace{0pt}
\includegraphics[width=0.49\textwidth]{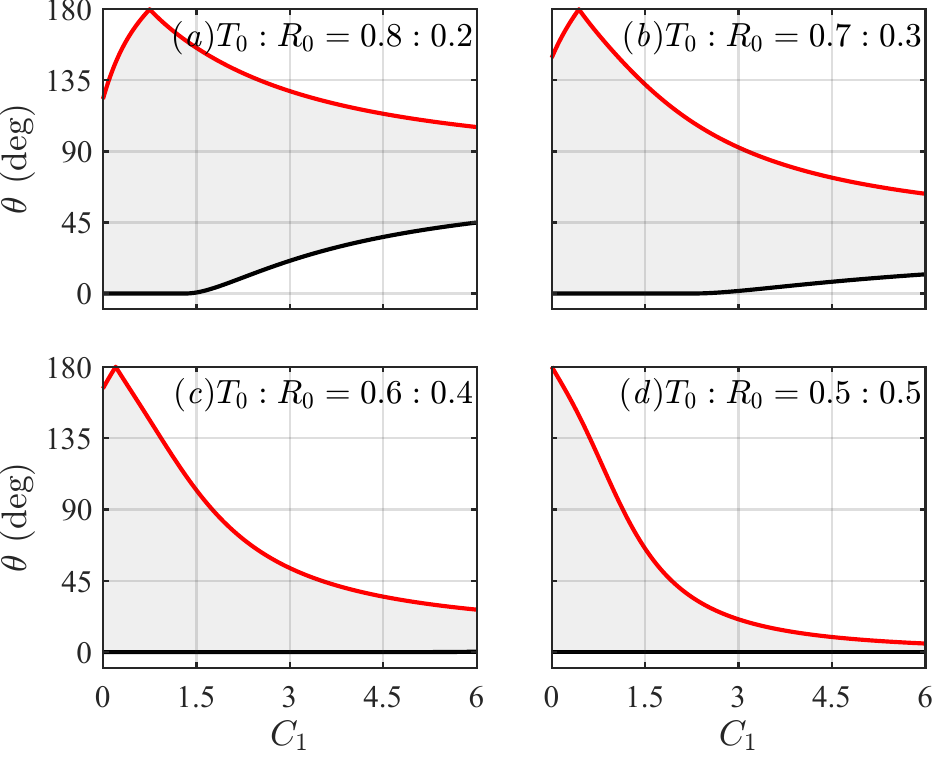}
\caption{Accessible range of the opening angle $\theta$ as a function of the initial intra-leg coupling strength $C_1$ for target probability partitions: ($a$) $T_0:R_0=0.2:0.8$, ($b$) $T_0:R_0=0.3:0.7$, ($c$) $T_0:R_0=0.4:0.6$ and ($d$) $T_0:R_0=0.5:0.5$. The red and black curves indicate the maximum and minimum accessible opening angles, respectively, with the accessible regions shaded in gray.}
\label{FIG3}
\end{figure}

\begin{equation}
T_0 = \frac{1}{2} \left[1+ \frac{
\alpha_1\alpha_2 +
\kappa^2\cos\!\left(\Delta\phi_{\kappa}\right)}
{\beta_1\beta_2}\right],
\label{eq:T_target}
\end{equation}
and $R_{0} = 1 - T_{0}$. Therefore the required phase change for the preset propagation angle $\theta_{0}$ and temporal reflection probability $R_{0}$ is
\begin{equation}
\Delta\phi_{\kappa}(\theta_0,R_0) =
\pm\arccos\!\left[
\frac{
\beta_1\beta_2(1-2R_0) -\alpha_1\alpha_2}
{\kappa^2} \right].
\label{eq:phase_target}
\end{equation}
Accordingly, the parameters after the temporal boundary are chosen as
\begin{equation}
C_2=C_1+\Delta C(\theta_0),\qquad
\phi_{\kappa 2} = \phi_{\kappa 1}
+ \Delta\phi_{\kappa}(\theta_0,R_0).
\label{eq:post_boundary_parameters}
\end{equation}

In Fig.~\ref{FIG2}, we present a numerical example to demonstrate the validity of our control strategy. As shown in Figs.~\ref{FIG2}($a_1$)--($a_3$), for the fine-tuned parameter sets, the separation between the two outgoing wave packets progressively increases. To show the opening angles more clearly, Figs.~\ref{FIG2}($b_1$)--($b_3$) present the trajectories of the 
\begin{figure}[htbp]
  \centering
    \vspace{0pt}
\includegraphics[width=0.48\textwidth]{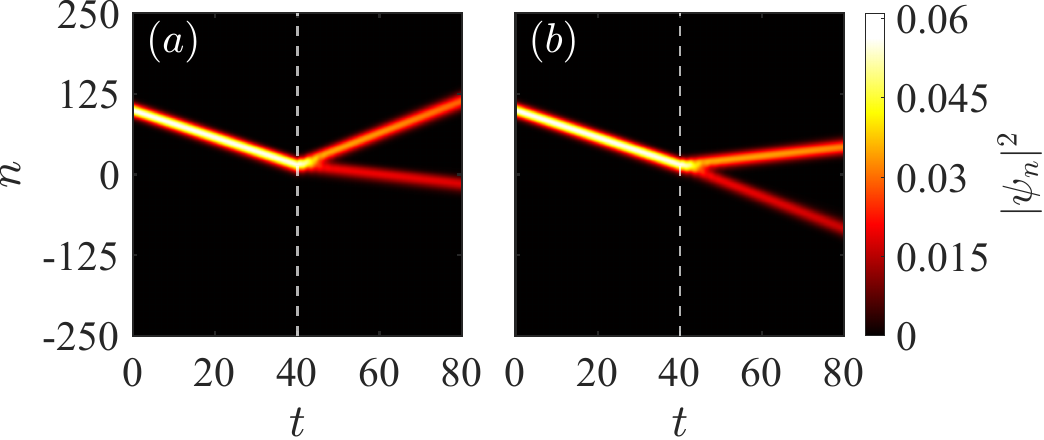}
\caption{Effect of reversing the post-boundary intra-leg
coupling phase on the propagation of the two outgoing
wave packets. ($a$) and ($b$) show the space--time intensity
$|\psi_n(t)|^2$ for $\phi_{C2}=+1.20\pi$ and
$\phi_{C2}=-1.20\pi$, respectively. The vertical dashed line marks the
temporal boundary at $t_c=40$. Here, $\phi_{\kappa}=0$, and all other parameters are the same as those in Fig.~\ref{FIG2}.}
\label{FIG4}
\end{figure}
wave-packet centers extracted from the peak positions of the intensity distributions. Here equal scales are used for the normalized
spatial and temporal coordinates. The opening angles extracted from the slopes of these trajectories are in good agreement with the preset values $60^\circ$, $90^\circ$, and $120^\circ$. In Figs.~\ref{FIG2}($c_1$)--($c_3$), we further show the wave-packet intensity profiles at $t_{end}=75$, when the two outgoing components are well separated. Different parameter sets may lead to slight differences in the wave-packet profiles and degrees of spreading, which leads to the small variations in the peak heights among the three cases. However, the probability partition, which is determined by the integrated intensities rather than the peak values, are all agree with the preset value $T_0:R_0=0.75:0.25$ within numerical accuracy. These results demonstrate that, with our control scheme, the geometry of the propagation can be well controlled while maintaining the temporal refraction-reflection probability partition, and vice versa.

Due to the compensating phase shift must be real in Eq. (\ref{eq:phase_target}), the range of the opening angle is therefore constrained. To demonstrate it, in Fig.~\ref{FIG3}, we present the accessible regions by the shaded areas for some typical probability ratios. For each partition, the upper bound varies nonmonotonically with $C_1$: it first increases toward the $180^\circ$ and then decreases. A nonzero minimum accessible opening angle, which increases with the imbalance between the reflection and refraction probabilities, also emerges at larger $C_1$. As the reflection probability increases from $0.2$ to $0.5$, the maximum accessible opening angle approaches $180^\circ$ shifts toward smaller $C_1$, while its value at large $C_1$ decreases substantially. These results demonstrate that the range of the opening angle depends on both the initial coupling strength and the target probability partition. The joint modulation scheme thus enables independent control within a well-defined accessible region, providing a quantitative criterion for selecting parameters that preserve the desired splitting probabilities.

We further investigate how the intra-leg coupling phase $\phi_{C}$ affects the propagation geometry while fixing the other parameters. It is interesting to note that when the sign of $\phi_{C}$ is reversed, the propagation angles of the temporal refraction and reflection components are exchanged, while their probability partition is preserved, as illustrated in Fig.~\ref{FIG4}. In both cases, the temporal
refraction and reflection probabilities are $T=0.3553$
and $R=0.6447$, and the propagation opening angle is
$\theta=104.75^\circ$. This behavior could be understood by noting Eq.~(\ref{eigen_energy}) that when $\phi_{C}$ changes to $-\phi_{C}$, the term $d_0(k)$  changes sign whereas $d_z(k)$ remains unchanged. Consequently, the eigenstates and hence the probabilities are unchanged, while the eigenenergies satisfy $E_{\pm}(k, -\phi_{C}) = -E_{\mp}(k, \phi_{C})$, leading to the group-velocity relation $v_{\pm} (k, -\phi_{C}) = -v_{\mp} (k, \phi_{C})$. Therefore, the two components exchange their propagation angles, while the opening angle remains unchanged.

\section{SSH Model}
\label{section3}
\begin{figure}[htbp]
\centering
\vspace{0pt}
\includegraphics[width=0.48\textwidth]{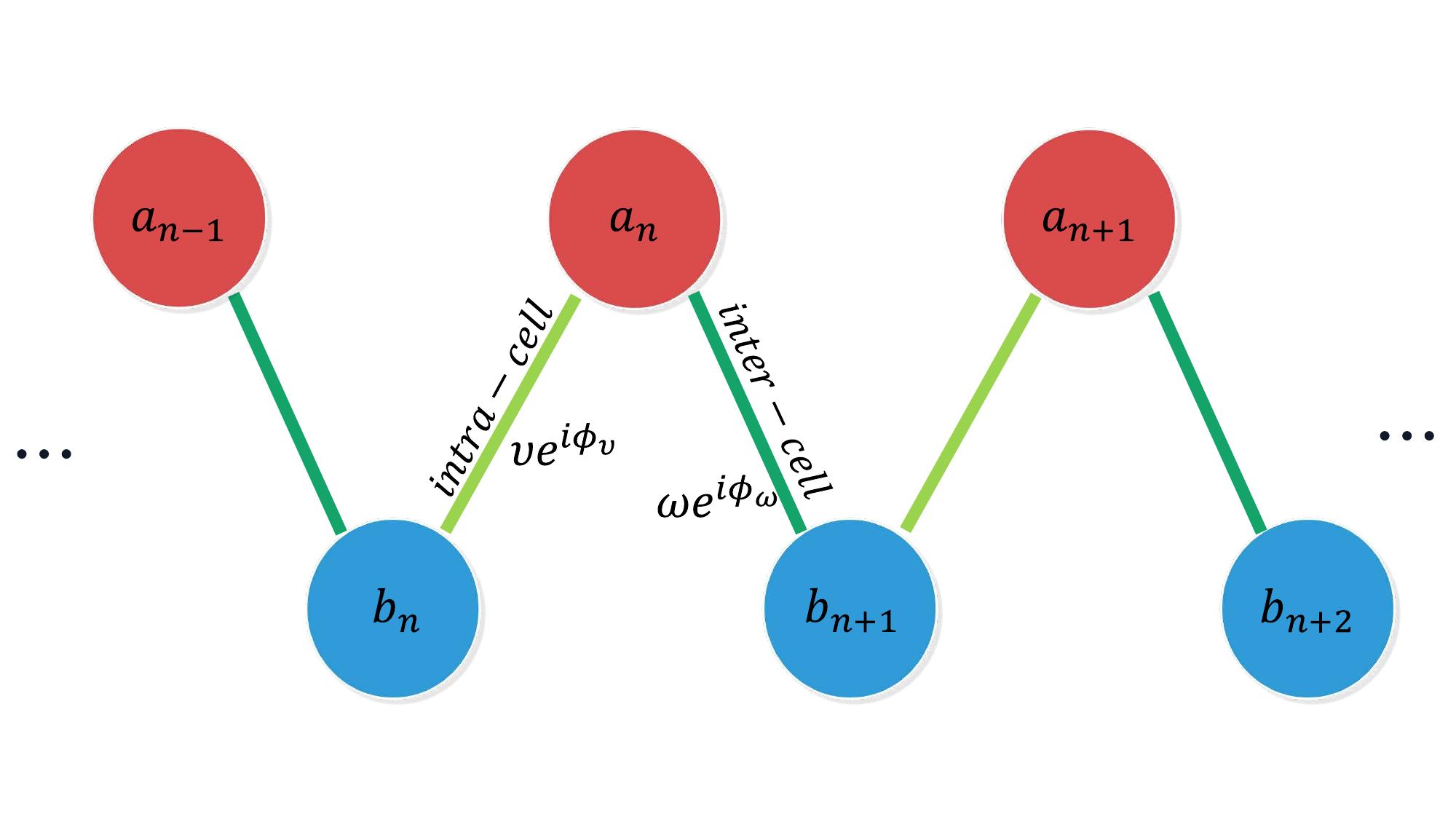}
\caption{Schematic of the SSH lattice with complex hopping
amplitudes. Red and blue circles denote sites of the $a$ and $b$ sublattices, respectively. The intra-cell hopping between $a_n$ and $b_n$ is $v e^{i\phi_v}$, while the inter-cell hopping between $a_n$ and $b_{n-1}$ is $\omega e^{i\phi_\omega}$. }
\label{FIG5}
\end{figure}
Now let's move to another representative system, the Su-Schrieffer-Heeger (SSH) model with temporal varying couplings, which has been experimentally realized in an ultracold-atom momentum lattice to study temporal boundary effect \cite{Dong2024}. In previous works \cite{Liu2026}, people have demonstrated that both the probability partition and the propagation angles can be controlled by tuning the coupling strengths or phases. However, varying the coupling strengths or the relative phase generally changes both the eigenstates and the dispersion relation, so that the probability partition and propagation angles generally change together. Here, we investigate how these quantities can be controlled independently by jointly tuning the coupling parameters. To introduce the scheme, we consider the SSH model illustrated in Fig.~\ref{FIG5}, such system can be described by the Hamiltonian 
\begin{equation}
H=\sum_n\left[v e^{i\phi_v}a_n^\dagger b_n+\omega e^{i\phi_\omega}a_n^\dagger b_{n+1}\right]+\mathrm{H.c.}，
\end{equation}

\begin{figure}[htbp]
\centering
\vspace{0pt}
\includegraphics[width=0.47\textwidth]{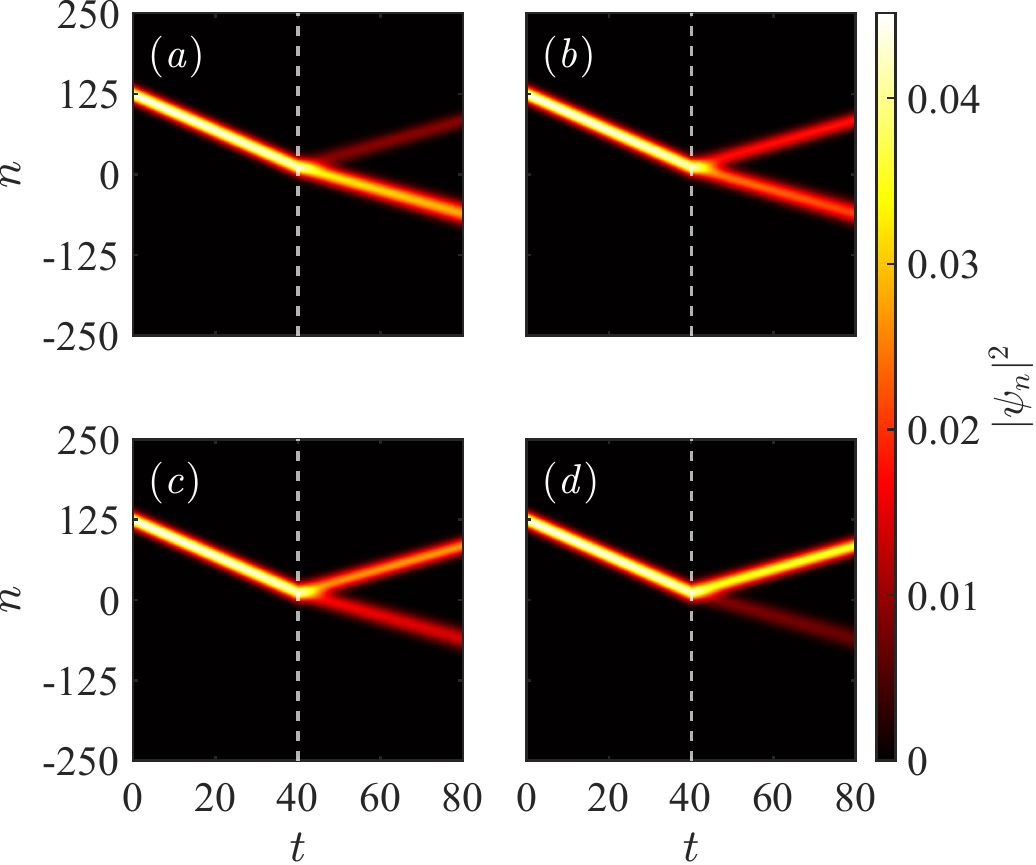}
\caption{Quantitative manipulation of temporal refraction--reflection probabilities by synchronously tuning $\phi_{v2}$ and $\phi_{\omega2}$. When the post-boundary phase difference $\Delta\phi_2=0.2\pi$, the space-time opening angle remains unchanged, while the energy distribution between the refraction and reflection branches can be continuously redistributed. The refraction-to-reflection probability ratios are $0.2:0.8$, $0.4:0.6$, $0.6:0.4$, and $0.8:0.2$ in panels (a)-(d), respectively. The parameters are set as  $k_0=\pi/2$, $v_1=w_1=v_2=w_2=4$, $\phi_{v1}=\phi_{w1}=0$.}
\label{FIG6}
\end{figure}
which takes the following form in momentum space
\begin{equation}
H(k)=\begin{pmatrix}0 &v e^{i\phi_v}+\omega e^{i\phi_\omega}e^{ik}\\v e^{-i\phi_v}+\omega e^{-i\phi_\omega}e^{-ik}&0\end{pmatrix}.
\end{equation}

The eigenenergies are
\begin{equation}
E_{\pm}(k)=\pm\sqrt{v^2+\omega^2
+2v\omega\cos\left(k+\phi_\omega-\phi_v\right)}.
\label{eq19}
\end{equation}
For $q(k)\neq 0$, the corresponding eigenstates can be written as
\begin{equation}
|\psi_{\pm}(k)\rangle
=\frac{1}{\sqrt{2}}
\begin{pmatrix}
\displaystyle\pm\frac{q(k)}{|q(k)|}\\
1
\end{pmatrix},
\label{eq20}
\end{equation}
where
\begin{equation}
q(k)=v e^{i\phi_v}+\omega e^{i\phi_\omega}e^{ik}.
\end{equation}

Assume the initial wave packet is prepared in the upper band and centered at $k_0$, its projections onto the post-boundary eigenstates give the temporal refraction and reflection probabilities
\begin{equation}
\begin{aligned}
T&=\frac{1}{2}\left[
1+\frac{\operatorname{Re}(q_1^*q_2)}{|q_1||q_2|}
\right],\\
R&=\frac{1}{2}\left[
1-\frac{\operatorname{Re}(q_1^*q_2)}{|q_1||q_2|}
\right].
\end{aligned}
\label{SSH_TR}
\end{equation}

\begin{figure*}[t]
\centering
\vspace{0pt}
\includegraphics[width=0.9\textwidth]{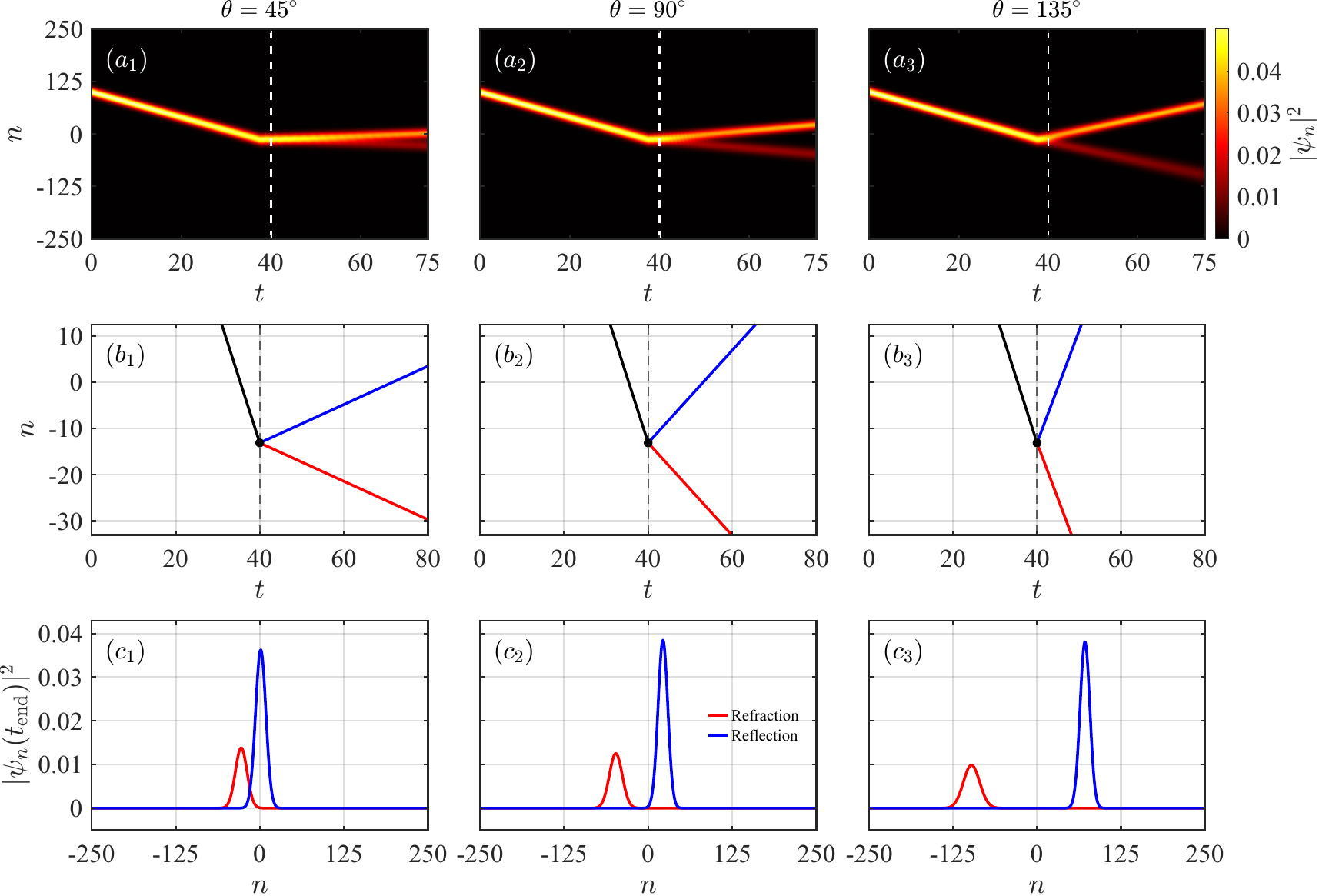}
\caption{Tunable space-time opening angle with fixed temporal refraction--reflection probability ratio in the SSH model. ($a_1$)--($a_3$) Intensity distributions obtained by synchronously rescaling the coupling strengths, where the opening angle is continuously tuned while maintaining the same energy partition $T_0:R_0=0.2:0.8$. ($b_1$)--($b_3$) Corresponding trajectories of the incident, temporally refracted, and temporally reflected wave packets, demonstrating the independent control of propagation geometry without modifying the refraction--reflection probability ratio. ($c_1$)--($c_3$) Spatial wave-packet intensity profiles  at $t_{end}=75$. The initial parameters as the same as Fig.~\ref{FIG6}.}
\label{FIG7}
\end{figure*}
%%%%%%%%%%%%%%%%%%%%%%%%%%%%%%%%%%%%%%%%
Here $q_j=q_j(k_0)$ is evaluated with the parameters before ($j=1$) or after ($j=2$) the temporal boundary. The corresponding group velocities follow from the eigenenergies
\begin{equation}
\begin{aligned}
v_{\pm}(k)
&=\frac{\partial E_{\pm}(k)}{\partial k}\\
&=\mp\frac{v\omega\sin\left(k+\phi_\omega-\phi_v\right)}
{\sqrt{v^2+\omega^2
+2v\omega\cos\left(k+\phi_\omega-\phi_v\right)}}.
\end{aligned}
\label{eq24}
\end{equation}

\begin{figure}[t]
\centering
\vspace{0pt}
\includegraphics[width=0.48\textwidth]{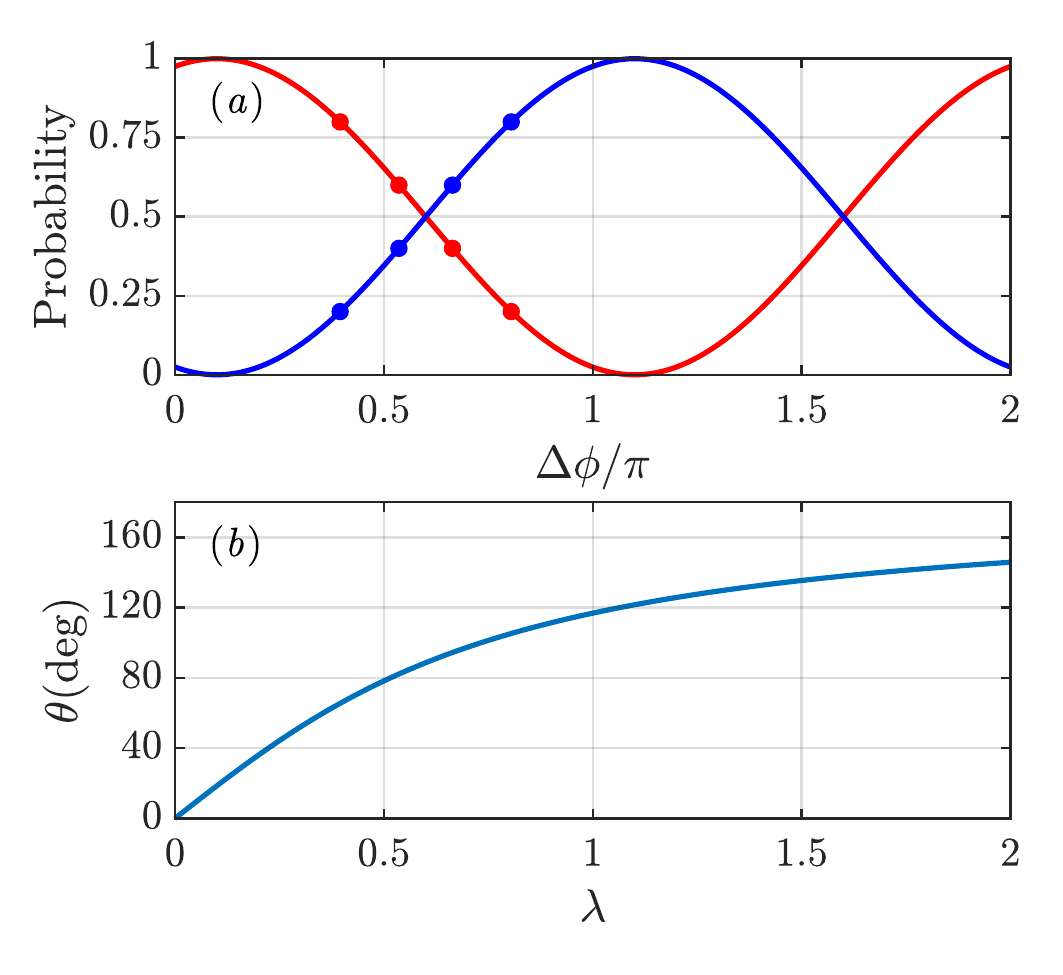}
\caption{Independent control of temporal refraction and reflection probabilities and the propagation opening angle in the complex-coupled SSH lattice. ($a$) Refraction probability $T$ (red) and reflection probability $R$ (blue) versus the coupling-phase shift $\Delta$$\phi$. Relevant parameters are the same as in FIG.~\ref{FIG6} and the corresponding dynamical evolution images of the marked points are shown in FIG. ~\ref{FIG6} ($a-d$). ($b$) Propagation opening angle $\theta$ versus the uniform coupling-strength scaling factor $\lambda$, with $v_1=3$, $w_1=4.5$, $v_2=3\lambda$, and $w_2=4.5\lambda$.}
\label{FIG8}
\end{figure}
%%%%%%%%%%%%%%%%%%%%%%%%%%%%%%%%%%%%%%%%%%%%%%%%
To achieve the independent control of the probability partition between temporal reflection and refraction branches at a fixed propagation opening angle, we synchronously apply a same phase shift $\Delta \phi$ on both the post-boundary coupling phases $\phi_{v}$ and $\phi_{\omega}$, leading to
\begin{equation}
    q_2\rightarrow e^{i\Delta\phi}q_2.
    \label{qk}
\end{equation}
According to Eq.~\eqref{eq19}, the eigenenergies, and hence the group velocities and propagation angles, depend on the phase difference $\phi_\omega-\phi_v$. Since the common phase shift leaves this difference unchanged, it preserves the propagation opening angles. Meanwhile, Eq.~\eqref{SSH_TR} gives
\begin{equation}
\begin{aligned}
T(\Delta\phi)
&=\cos^2\left[\frac{\Delta\phi+\Delta\vartheta_0}{2}\right],\\
R(\Delta\phi)
&=\sin^2\left[\frac{\Delta\phi+\Delta\vartheta_0}{2}\right],
\end{aligned}
\end{equation}
where $\Delta\vartheta_0=\arg q_2(k_0)-\arg q_1(k_0)$ is the phase difference between the post- and pre-boundary coupling functions before introducing $\Delta\phi$. Therefore, the common hopping-phase shift continuously tunes the temporal refraction and reflection probabilities while preserving the propagation opening angle. As illustrated in Fig. \ref{FIG6}$(a)-(d)$, the preset probability ratio $T_0:R_0$ changes from $0.2:0.8$ to $0.8:0.2$, whereas the trajectories of the two outgoing branches retain the same slopes. %需要解释吗？Although the local peak brightness is not strictly symmetric after swapping energy ratios due to different wavepacket broadening, the total integrated probability of each branch still satisfies the preset ratio.%

We further investigate how to tune the propagation angles while maintaining the fixed probability partition. To achieve it, a scaling factor $\lambda$ is simultaneously applied onto both the post-boundary inter- and intra-cell coupling strengths $v$ and $\omega$. Note that the eigenstates depend on the ratio $q_2(k)/|q_2(k)|$, therefore the introduction of such scaling factor leaves the projections onto the two post-boundary bands, i.e., the  probability partition unchanged. Meanwhile, the eigenenergies and hence the group velocities would be scaled by $\lambda$, result in the opening angle changing as
\begin{equation}
\begin{aligned}
\theta(\lambda)
&=2\arctan\!\left[
\frac{\lambda|v_2\omega_2\sin\delta_2|}
{\sqrt{v_2^2+\omega_2^2+2v_2\omega_2\cos\delta_2}}
\right],
\end{aligned}
\label{eq:ssh_angle_scaling}
\end{equation}
where, $\delta_2=k_0+\phi_{\omega2}-\phi_{v2}$. Fig.~\ref{FIG7} illustrates the tuning of the propagation opening angle at a fixed probability partition $T_0:R_0=0.2:0.8$. As shown in the upper panels ($a_1$)–($a_3$), increasing the common scaling factor $\lambda$ increases the magnitudes of the post-interface group velocities, resulting in faster separation of the temporally refracted and reflected wave packets. The corresponding trajectories in panels ($b_1$)–($b_3$) yield the prescribed opening angles $\theta=45^\circ$, $90^\circ$ and $135^\circ$, respectively. Panels ($c_1$)–($c_3$) show the spatial intensity profiles of the refracted (red) and reflected (blue) components at the same time, further illustrating the increasing separation and the preserved intensity distribution. Although the peak heights and widths slightly vary among the three cases, the integrated probabilities are strictly fixed. These results confirm that the opening angle can be tuned independently of the probability partition in the SSH model.
Finally, we present the accessible tuning ranges of the probability partition and propagation opening angle in Fig.~\ref{FIG8}. In the upper panel ($a$), varying the common phase shift $\Delta\phi$ from $0$ to $2\pi$ at a fixed opening angle allows $T$ and $R$ to vary complementarily over the full interval $[0,1]$, consistent with the $\cos^2$ and $\sin^2$ dependences derived above. Complete temporal reflection or refraction can therefore be achieved by selecting an appropriate phase shift. In panel ($b$), with the probability partition held fixed, the opening angle increases monotonically with $\lambda$ at a progressively decreasing rate, approaching $180^\circ$ in the large-$\lambda$ limit, as expected from its arctangent dependence. These results demonstrate that the probability partition and opening angle can be preset independently within their respective continuous tuning ranges. This flexibility provides a theoretical basis for programmable temporal beam splitting and wave routing in the SSH model, with separately adjustable splitting ratios and propagation geometry.
%%%%%%%%%%%%%%%%%%%%%%%%%%%%%%%%%%%%%%%%%%%%%%%%%
\section{Conclusions}
In this work, we have demonstrated independent control of the temporal reflection and refraction probabilities and the propagation opening angle in two-leg ladder and complex-coupled SSH models. In the ladder model, coupling-strength modulation sets the opening angle, while a compensating change in the interleg coupling phase preserves the prescribed probability partition. Reversing the intraleg coupling phase further reverses the common drift direction without changing either the splitting probabilities or the opening angle. In the SSH model, a common shift of the hopping phases controls the probability partition at fixed dispersion, whereas a uniform rescaling of the hopping strengths tunes the opening angle without modifying the probability partition. Analytical control relations and numerical wave-packet simulations establish these results within the accessible parameter ranges. Our results provide a theoretical basis for programmable time-modulated optical devices. The resulting temporal scattering elements provide building blocks for beam splitters, routers, and temporal grating analogues based on controlled sequences of temporal interfaces.

\section*{Acknowledgments}
This work is supported by the National Natural Science
Foundation of China (Grants No. 12564048,
No. 12564047), the Hainan Provincial Natural Science Foundation of China (Grant No. 122QN302) and Hainan
Provincial Graduate Innovative Research Project (Grant No.
Hys2025-379).
\twocolumngrid
\appendix
\renewcommand{\appendixname}{\MakeUppercase{APPENDIX}}
\end{CJK*}
\bibliographystyle{apsrev4-2}
\bibliography{sample}
\end{document}